\documentclass[aip,graphicx]{revtex4-1}

\usepackage{graphicx}
\usepackage{dcolumn}
\usepackage{bm}
\usepackage{bbold}
\usepackage{subfig}
\usepackage{siunitx}
\usepackage[utf8]{inputenc}
\usepackage[T1]{fontenc}
\usepackage{mathptmx}
\usepackage{etoolbox}

\makeatletter
\def\@email#1#2{%
 \endgroup
 \patchcmd{\titleblock@produce}
  {\frontmatter@RRAPformat}
  {\frontmatter@RRAPformat{\produce@RRAP{*#1\href{mailto:#2}{#2}}}\frontmatter@RRAPformat}
  {}{}
}%
\makeatother
\begin{document}


\title[]{Electrosorption-induced deformation of a porous electrode with non-convex pore geometry in electrolyte solutions: a theoretical study}

\author{Andrei L. Kolesnikov}
\email{kolesnikov@inc.uni-leipzig.de}
\affiliation{Institut f\"ur Nichtklassische Chemie e.V., Permoserstr. 15, 04318 Leipzig, Germany}
\author{Daria A. Mazur}
\affiliation{School of Applied Mathematics, HSE University, Tallinskaya st. 34, 123458 Moscow, Russia}
\author{Yury A. Budkov}
\email{ybudkov@hse.ru}
\affiliation{School of Applied Mathematics, HSE University, Tallinskaya st. 34, 123458 Moscow, Russia}
\affiliation{G.A. Krestov Institute of Solution Chemistry of the Russian Academy of Sciences, 153045, Akademicheskaya st. 1, Ivanovo, Russia}

\begin{abstract}
Porous carbon is well known as a good candidate for the development of electrochemical double-layer capacitors. Predominantly, many conventional carbons are microporous and often well described by the assumption of slit pore geometry. However, there is a class of carbons that is significantly different from the others, namely templated mesoporous carbons. In this work, we study electrosorption-induced deformation in CMK-3-like mesopores having nonconvex geometry applying a mean-field approach. The model is based on the modified Poisson-Boltzmann equation taking into account the excluded volume of the ions within the hard-sphere model in the Percus-Yevick approximation. We assume that the deformation is caused by two effects: ion osmotic pressure and electrostatic interactions of the electric double layers on charged rods. The latter were calculated via the Maxwell stress tensor agreeing with the modified Poisson-Boltzmann equation. We estimated the pore-load modulus of the CMK-3-like material and found an agreement with the previously obtained values by small-angle neutron scattering (SANS). Additionally, we studied the differential capacitance in the non-convex pore geometry and found that the behavior of the differential capacitance profiles was similar to that of the profiles obtained for flat electric double layers. Namely, we obtained the crowding regime at rather high voltages and more pronounced profile asymmetry with increasing differences in the ionic sizes.   
\end{abstract}
\maketitle
\section{Introduction}
Porous carbon is well known as a good candidate for the development of electric double layer (EDL) capacitors. More precisely, it was chosen due to its large specific surface area, electrochemical stability, and conductivity~\cite{shao2020nanoporous}. The storage mechanism is similar to classical physisorption, i.e. the charge is stored on the electrode/electrolyte interface. Of course, these are not the only similarities and both adsorption~\cite{kolesnikov2021models,gor2017adsorption} and electrosorption~\cite{augustyn2020deformation,hantel2014parameters} are accompanied by the host material deformation. Basically, an application involving a large number number of charging/discharging cycles can cause degradation of the porous electrode due to the repeating deformations. That may cause capacitor failure. However, these strains were not always undesirable and can be used, for instance, in the designing of voltage-driven actuators~\cite{torop2011flexible}. Though electrosorption-induced deformation is a long-known phenomenon, its mechanism is not fully understood and depends not only on the electrolyte and solvent but also on the adsorbent. For instance, pore size distribution or magnitude of surface area can affect the resulting deformation. In this context, CMK-3 templated carbon~\cite{jun2000synthesis} or monolithic templated carbon with a similar mesoporosity structure~\cite{putz2020hierarchically} are a very interesting example due to a peculiar structure and pore geometry. Both of them have nonconvex pores formed by the voids between hexagonally packed carbon cylindrical rods, and, thus, differ a lot from typical microporous carbons. 

In Ref.\cite{koczwara2017situ} the electrosorption-induced deformation of carbon-based supercapacitor electrodes with CMK-3-like mesoporosity was measured by means of in situ small-angle X-ray scattering. The obtained strains were positive and asymmetric, demonstrating higher values at negative voltages. The authors~\cite{koczwara2017situ} proposed a deformation mechanism consisting of two major contributions: ion osmotic pressure and C-C bond length variation due to the electron/hole doping during the electrode charging. The interest in these particular measurements is warmed up by the existence of experimentally measured adsorption-induced deformation in the very similar materials~\cite{ludescher2021adsorption} and estimated pore-load moduli. Thus, the existence of a correct electrosorption model would allow cross-checking the pore-load moduli and possibly shed some light on the deformation mechanism of other materials with similar pore geometry. 

The adsorption-induced deformation of the CMK-3-like material has been already theoretically studied by two of us\cite{kolesnikov2020density}. The deformation was considered solely originated from the change in the inter-rod distances. We assumed that each rod was connected with the nearest neighbors by virtual springs with the elastic constant $\kappa$. The springs were distributed along the rod axis with the density $\rho_{spr}$. In the present communication, we use the proposed model~\cite{kolesnikov2020density} of a CMK-3-like material to describe the electrosorption-induced deformation. We utilize the modified Poisson-Boltzmann equation with the corresponding stress tensor~\cite{budkov2022} to calculate the distribution of the resultant forces acting on the carbon rods. In turn, this allows us to additionally check the previously proposed deformation mechanism and estimate the pore-load modulus by the classical mean-field theory.

\section{Model}

\begin{figure}[ht]
\centering
\includegraphics[width=0.5\linewidth]{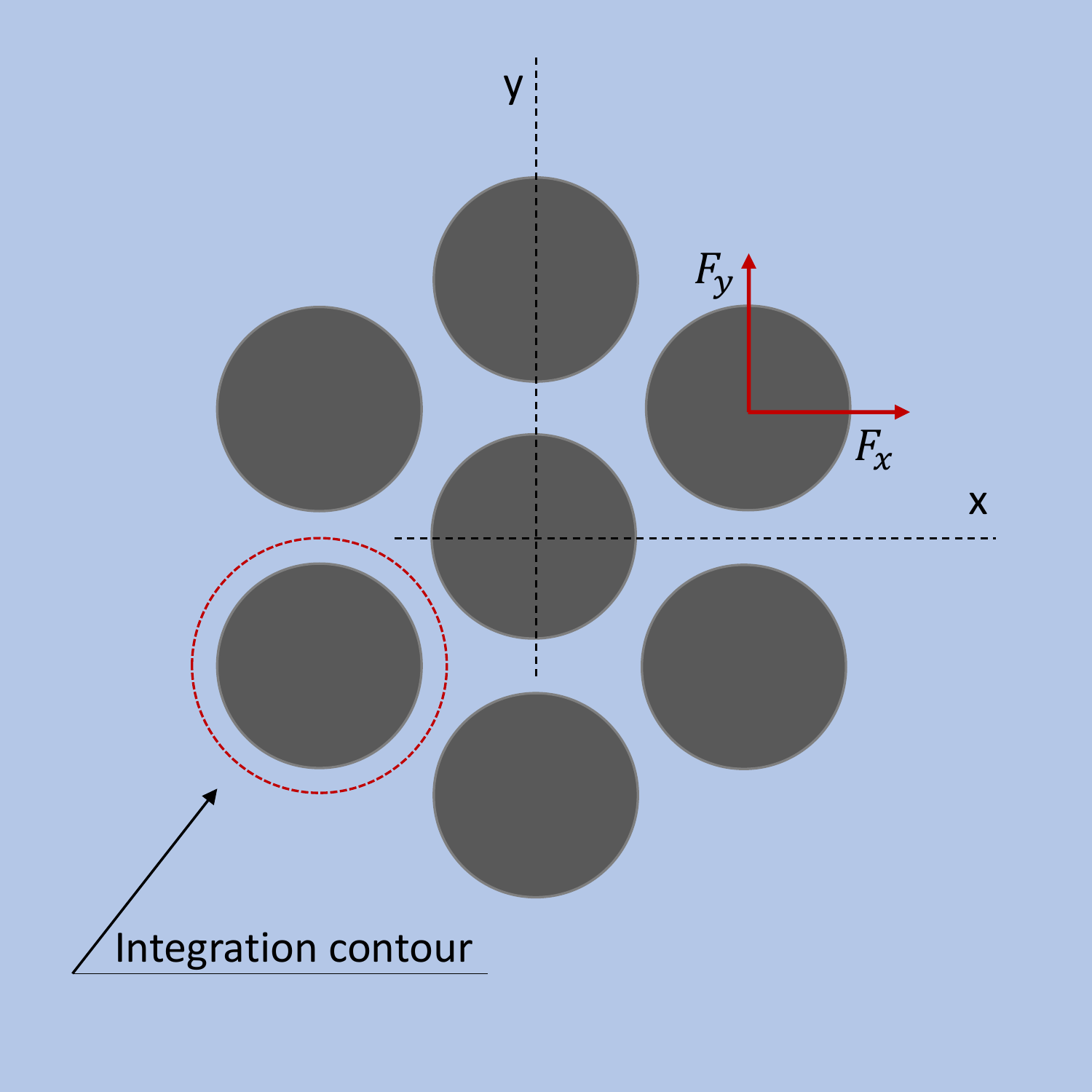}
\caption{ Schematic representation of the system under consideration. The dotted lines represent the x/y-axis, the red arrows are the two projections of the force acting on the carbon rod and the red dotted circle is a possible projection of the integration contour. }
\label{fig:system} 
\end{figure}

The system geometry aiming to describe electrolyte-mediated rod-rod interactions in CMK-3-like material is presented in Fig.\ref{fig:system}. It consists of seven rods of a certain radius, $R$, placed in the centers of hexagonal tiling. The free space between the rods is filled with a 1:1 aqueous electrolyte solution. As CMK-3 is an inverse replica of SBA-15, the rods are interconnected via thin carbon bridges, which are not taken into account explicitly in the model. However, their extension is one of the possible sources of mesoporous adsorption/electrosorption-induced deformation observed in such types of materials~\cite{koczwara2017situ,ludescher2021adsorption}. The previously proposed model of CMK-3 deformation~\cite{kolesnikov2020density} is utilized in the current study. Additionally, we neglect the effect of solvent perturbations due to the existence of mobile ions in the solution. The total force acting of each rod is a sum of two contributions: electrostatic and steric. The latter is the osmotic pressure of the mobile ions and the former is the electrostatic force acting on the surface charge~\cite{neu1999wall,trizac1999long}. Both of them can be obtained within the framework of the modified Poisson-Boltzmann equation\cite{budkov2022}, which can be summarized as the following system of equations:
\begin{eqnarray}
\varepsilon\mathbf{\nabla}^2\psi &=& -q(c_{\rm +} - c_{\rm -}) \label{eq:system}\\
\mu_{\rm id}(c_{\rm \pm}) + \mu_{\rm hs}(c_{\rm \pm}) \pm q\psi &=& \mu_{\rm 0}(c_{\rm 0, \pm}) \nonumber 
\end{eqnarray}
where $\varepsilon$ is the dielectric permittivity of the solution, $\psi$ is the local electrostatic potential and $\/c_{\rm \pm}$ are the number densities (concentrations) of the mobile ions in the solution; $\mu_{\rm id}=k_{B}T\ln\left(c_{\pm}\Lambda_{\pm}^3\right)$ and $\mu_{\rm hs}$ are the ideal and hard-spherical parts of the total chemical potential ($\Lambda_{\pm}$ are the de Broglie thermal wavelengths of the ions), $\mu_{\rm 0}$ is the total chemical potentials in the bulk solution. The force acting on each rod is a surface integral of the total stress tensor ($\boldsymbol{\sigma}$):
\begin{equation}
    \mathbf{F} = \oint_{\rm S} \boldsymbol{\sigma} \cdot \mathbf{n} dS,
\end{equation}
where $\mathbf{n}$ is the external normal to the integration surface (which is a cylinder in our case) and the stress tensor is
\begin{equation}
    \boldsymbol{\sigma} = \varepsilon \mathbf{\nabla} \psi \otimes \mathbf{\nabla} \psi - \biggl[P(c_{\rm +}, c_{\rm -}, T) + \frac{\varepsilon(\mathbf{\nabla}\psi)^{\rm 2}}{2}\biggl]\mathbb{1},
\end{equation}
where $P(c_{+},c_{-},T)$ is the osmotic pressure of the ions and $\mathbb{1}$ is the unit second order tensor.
The corresponding volume density force has the following form
\begin{equation}
    \mathbf{f} = -\mathbf{\nabla} P(c_{+},c_{-}, T) - q(c_{\rm +} - c_{\rm -})\mathbf{\nabla}\psi,
\end{equation}
which leads to the standard condition of mechanical equilibrium $\mathbf{\nabla} P(c_{+},c_{-}, T) = - q(c_{\rm +} - c_{\rm -})\mathbf{\nabla}\psi$ \cite{hansen1990theory}, because the tensor is divergence free\cite{neu1999wall}. Since each integration surface has cylindrical geometry, we can write down the force projections on the x/y-axis in the following way:
\begin{equation}
    F_{\rm x} = LR_c\int\limits_0^{\rm 2\pi} d\varphi  \biggl\{-\biggl[ P + \frac{\varepsilon E^2}{2}\biggl]\cos\varphi + \varepsilon E_{\rm x} (E_{\rm x}\cos\varphi + E_{\rm y}\sin\varphi)\biggl\} \label{eq.Fx}
\end{equation}
and
\begin{equation}
    F_{\rm y} = LR_c\int\limits_0^{\rm 2\pi} d\varphi  \biggl\{-\biggl[ P + \frac{\varepsilon E^2}{2}\biggl]\sin\varphi + \varepsilon E_{\rm y} (E_{\rm x}\cos\varphi + E_{\rm y}\sin\varphi)\biggl\}, \label{eq.Fy}
\end{equation}
where $R_c$ is the radius of an integration contour around the sole carbon rode and $\mathbf{E}$ is the electric field at a certain point with its components ($E_{\rm x}$, $E_{\rm y}$). The integration surface does not necessarily coincide with the geometrical cylinder surface because the tensor is divergence free. However, it must always include only one rod. In case when the integration is made precisely over the rod surface the osmotic pressure does not contribute to the force and can be omitted in Eqs. (\ref{eq.Fx} and \ref{eq.Fy}).
It was shown that CMK-3-like material adsorption-induced strain also follows the logarithmic dependence on relative pressure at small under-saturation in the bulk vapor phase~\cite{kolesnikov2020density,ludescher2021adsorption}:
\begin{equation}
    \epsilon = \frac{1}{M} (P_{\rm v, 0} - P_{\rm v} + k_{\rm B} T \rho \ln(P_{\rm v}/P_{\rm v, 0})),
\end{equation}
where $M$ is the pore-load modulus equal to $M = \sqrt{3}\rho_{\rm spr} \kappa$, $k_{\rm B}$ is the Boltzmann constant, $T$ is the temperature, $P_{\rm v}$ is the vapor pressure during an adsorption experiment and $P_{\rm v, 0}$ is the saturated vapor pressure. Using the same deformation mechanism, which was described above, we can express the electrosorption-induced strain via the forces acting on the rods and pore-load modulus:
\begin{equation}
    \epsilon_{\rm el} = \frac{\sqrt{3}f_{\rm el}}{2 d M} \label{eq:M},
\end{equation}
where $f_{\rm el}$ and $\epsilon_{\rm el}$ are the length density of the generated force and the corresponding deformation, respectively, and $d$ is the inter-rod distance. Thus, using this equation, we can estimate the consistency of both models.

\begin{figure}[ht]
\subfloat[2R = \SI{6}{nm}]{\includegraphics[width=.3\textwidth]{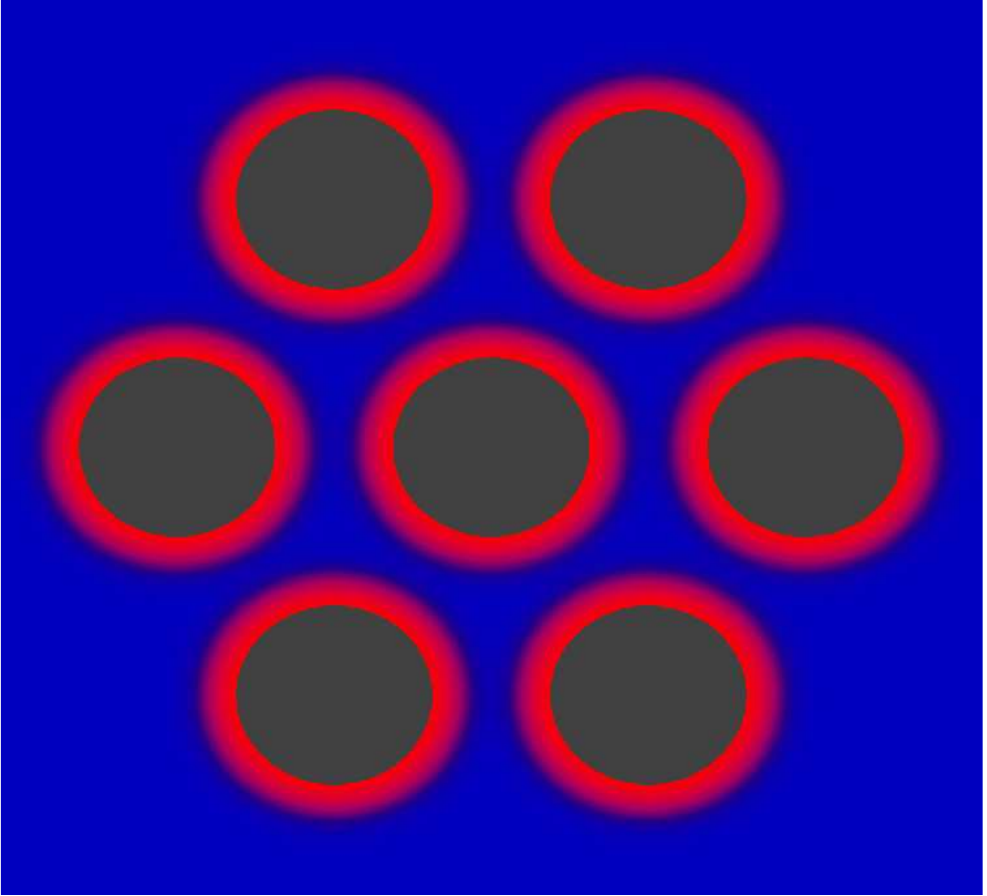} \label{fig:cl_6_06}  } 
\subfloat[2R = \SI{6.5}{nm}]{\includegraphics[width=.3\textwidth]{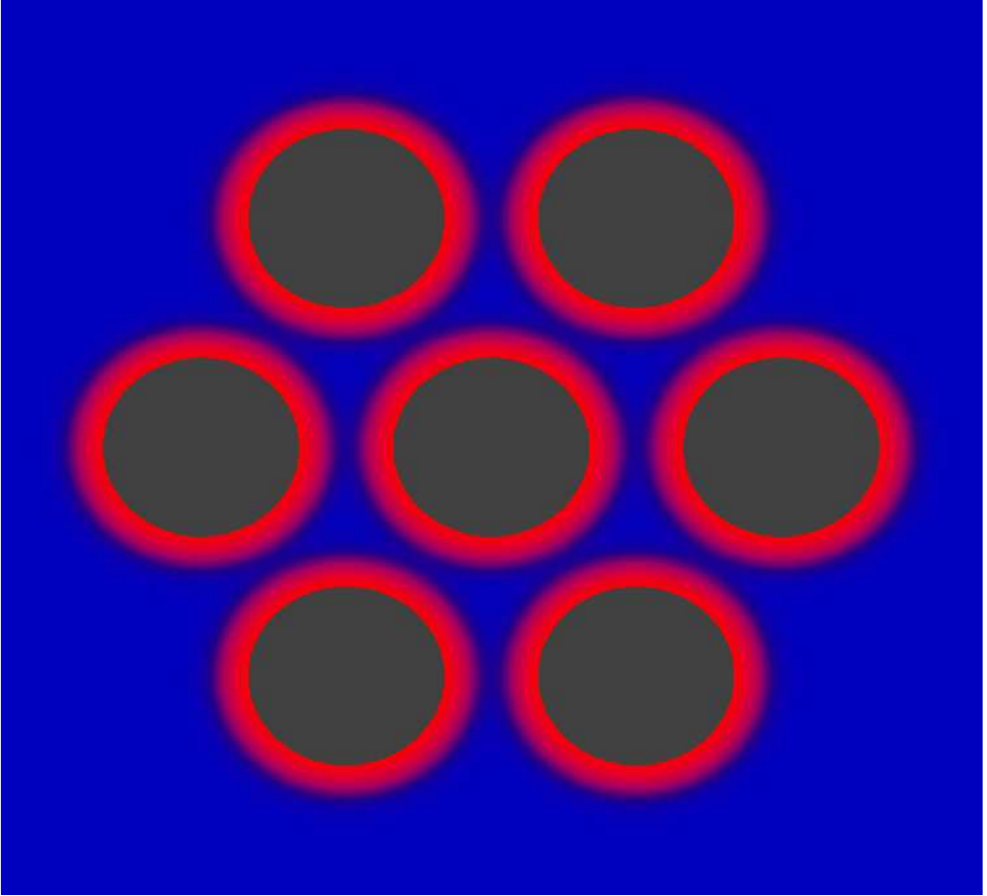} \label{fig:cl_65_06} }
\subfloat[2R = \SI{7}{nm}]{\includegraphics[width=.355\textwidth]{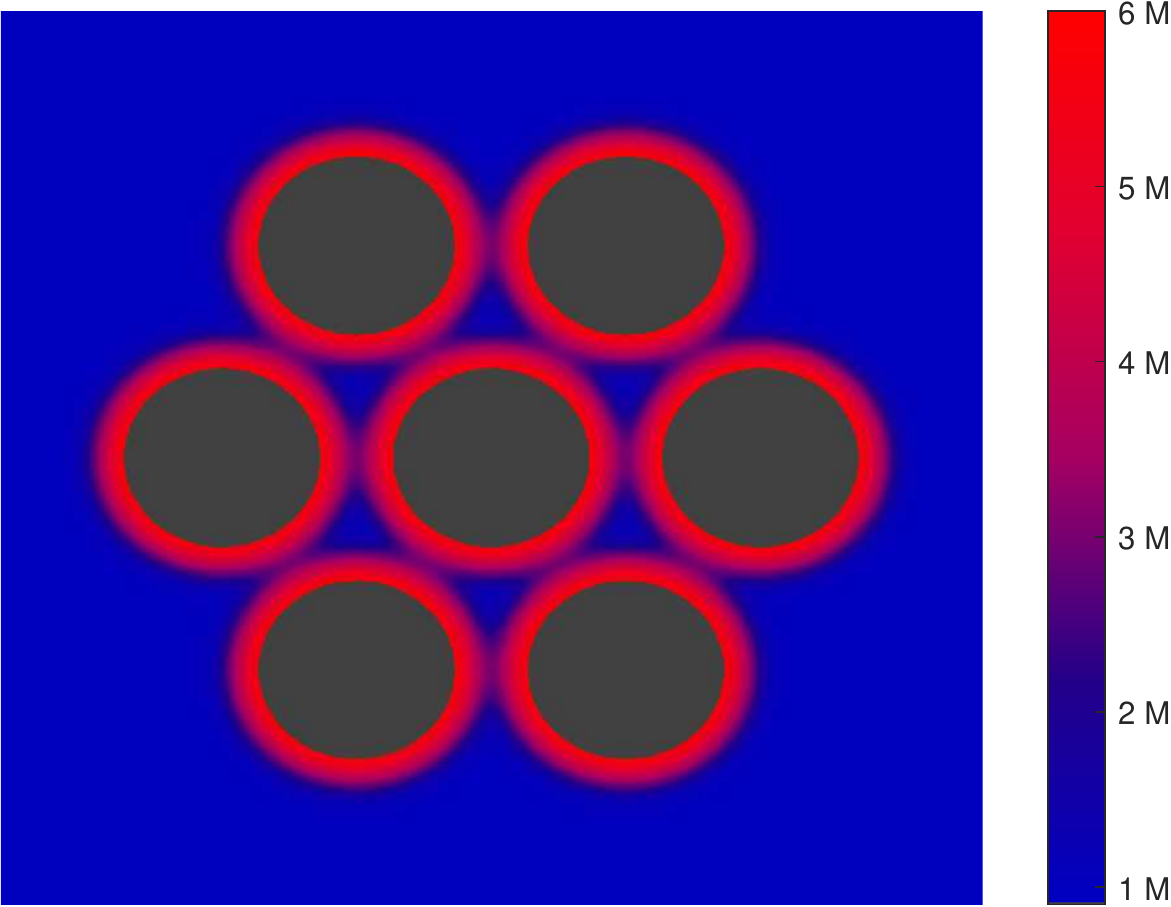} \label{fig:cl_7_06}}\\
\subfloat[2R = \SI{6}{nm}]{\includegraphics[width=.3\textwidth]{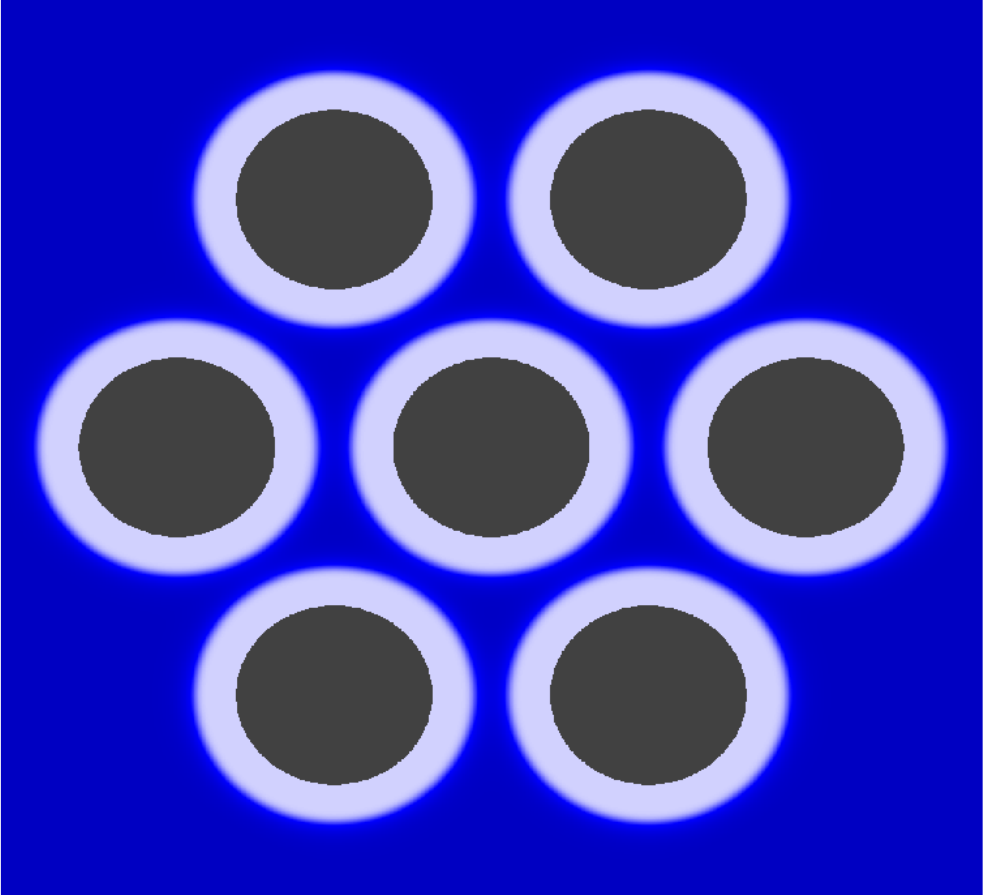} \label{fig:li_6_06}}
\subfloat[2R = \SI{6.5}{nm}]{\includegraphics[width=.3\textwidth]{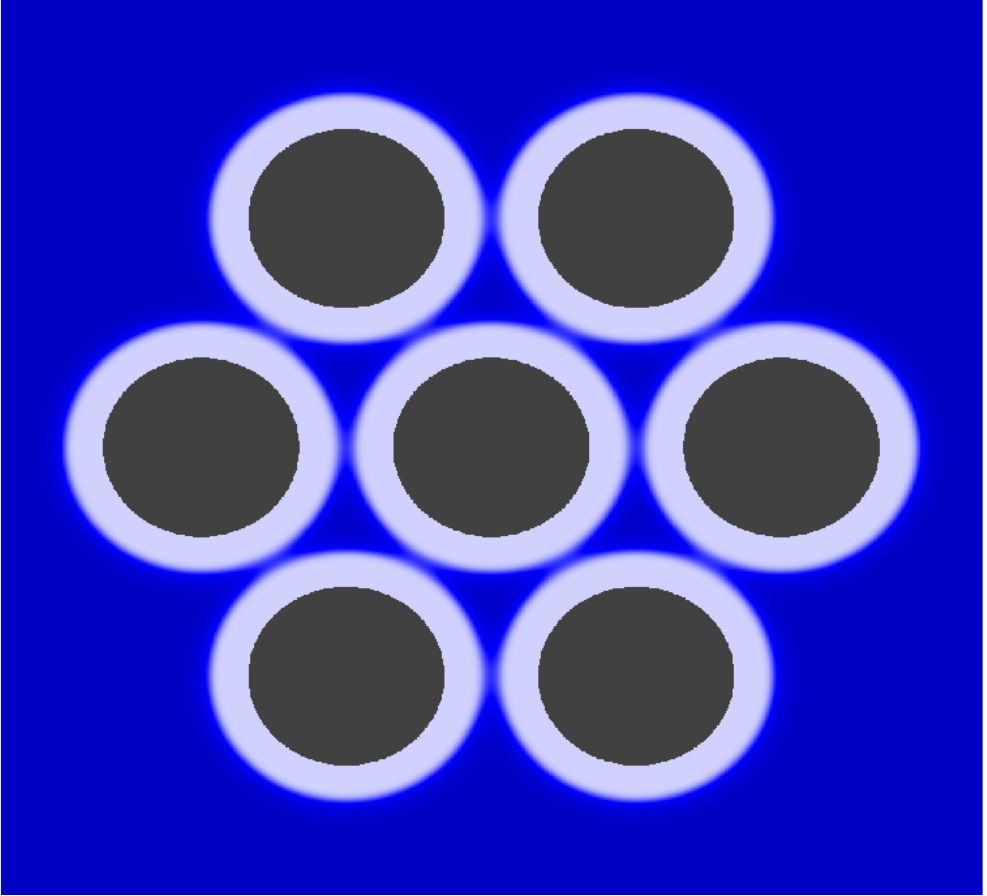} \label{fig:li_65_06}}
\subfloat[2R = \SI{7}{nm}]{\includegraphics[width=.365\textwidth]{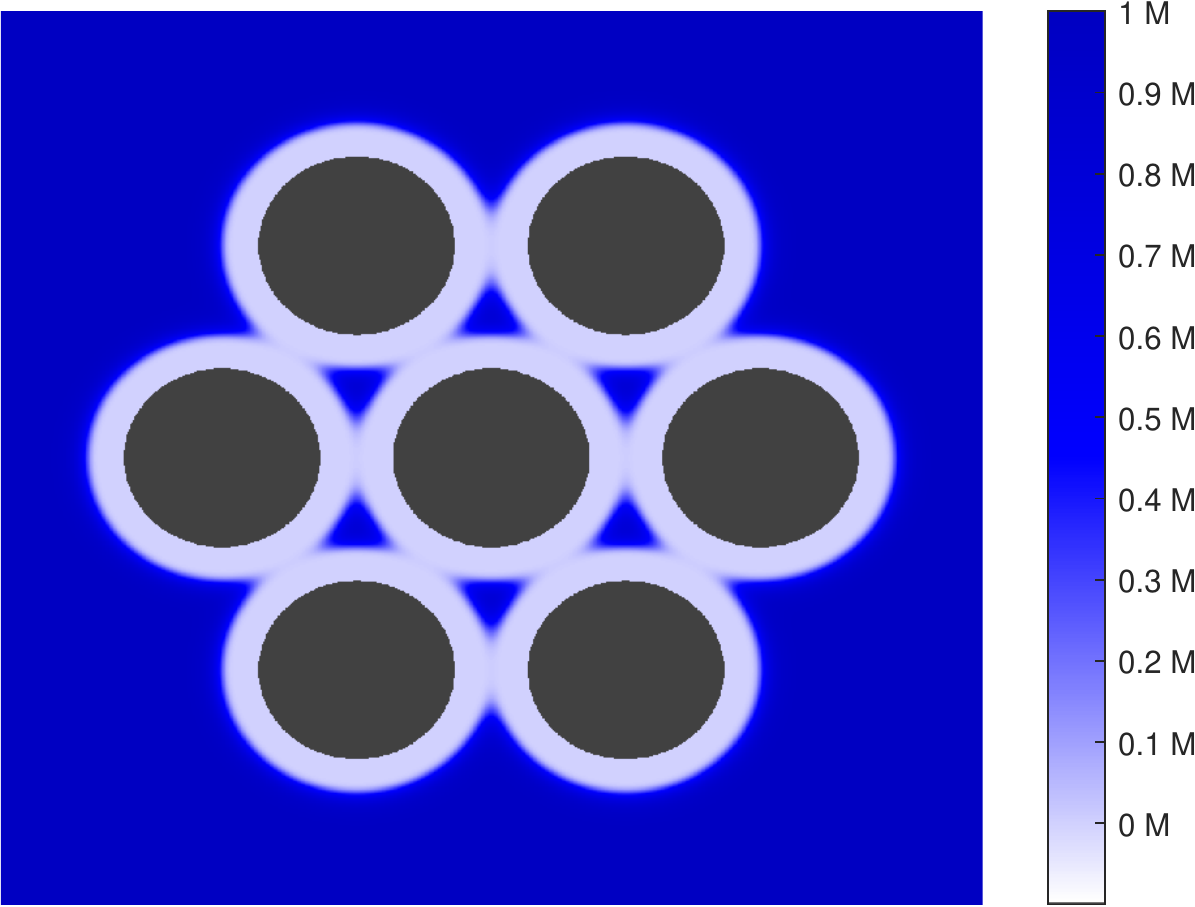} \label{fig:li_7_06}}
\caption{Anion (upper row) and cation (bottom row) number densities for three different rod diameters (2R = \SI{6}{nm}, \SI{6.5}{nm} and \SI{7}{nm}) at the carbon voltage $V_{\rm c}$ = \SI{0.6}{\volt} for the LiCl electrolyte solution.}
\label{fig:densities}
\end{figure}

\section{Results and discussion}
We solved the system of equations (\ref{eq:system}) for three different rod diameters, namely \SI{6}{nm}, \SI{6.5}{nm} and \SI{7}{nm}. These values were reported in Ref.\cite{koczwara2017situ} along with the interrod distance d $\approx$ \SI{9.6}{nm}. Fig.\ref{fig:densities} demonstrates the anion and cation density distributions for the aqueous LiCl electrolyte solution. The calculations were performed at the fixed carbon voltage $V_{\rm c}$ = \SI{0.6}{\volt}. There was a positive excess of anions and a negative excess of cations in the vicinity of the rods due to the positive carbon voltage. The case of \SI{6.5}{nm} rod diameter is the boundary between the two regimes obtained: without (Figs.(\ref{fig:cl_6_06}) and (\ref{fig:li_6_06})) and with (Figs.(\ref{fig:cl_7_06}) and (\ref{fig:li_7_06})) pronounced EDL overlapping. The first regime is also similar to the ion distributions at lower carbon voltages and differs mostly by the EDL width. The second regime is only observed at high carbon voltages and is characterized by almost total depletion of cations from the interior space.  
\begin{figure}[ht]
\subfloat[LiCl]{\includegraphics[width=.51\textwidth]{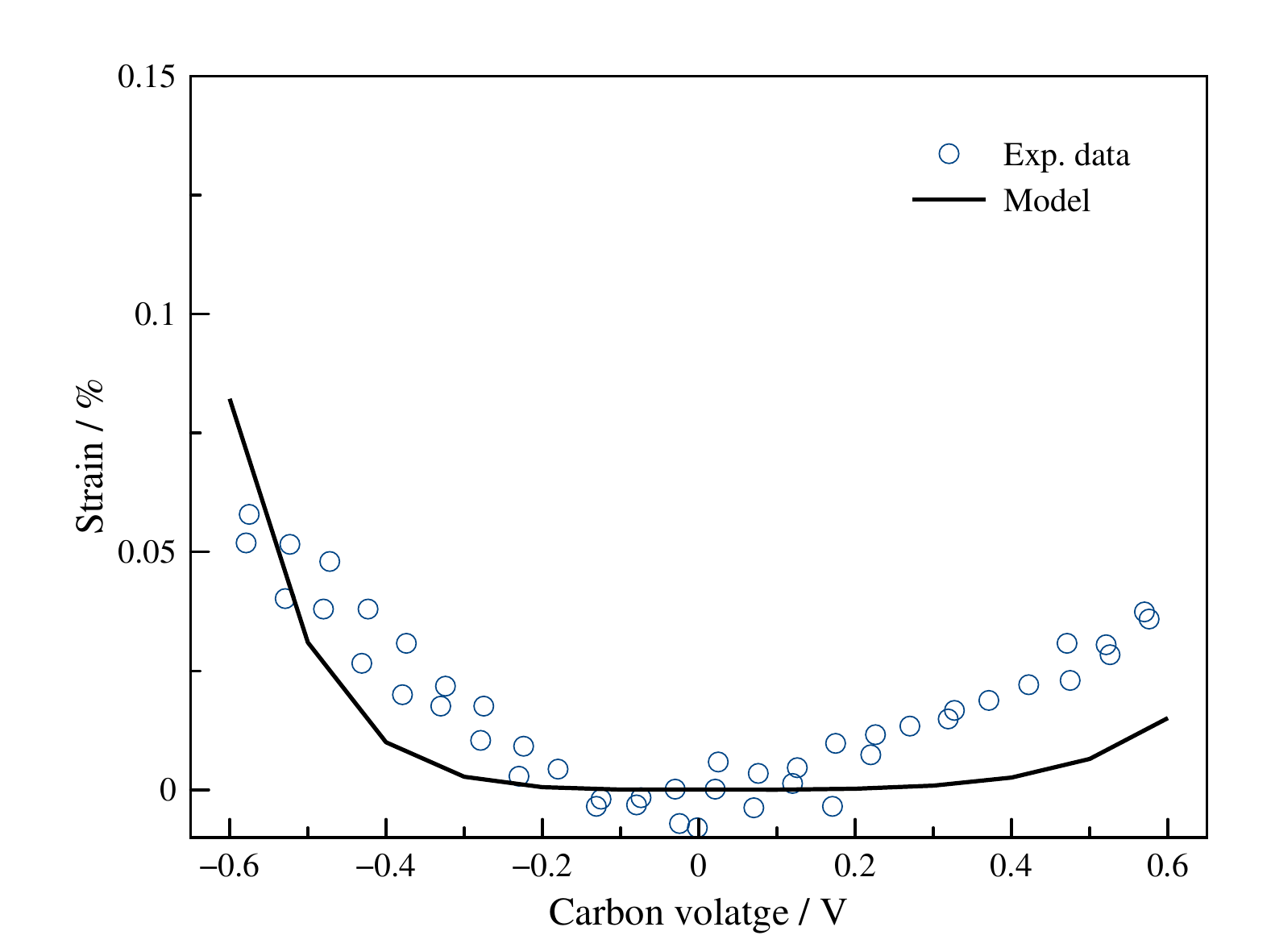} \label{fig:licl_fit}  } 
\subfloat[CsCl]{\includegraphics[width=.51\textwidth]{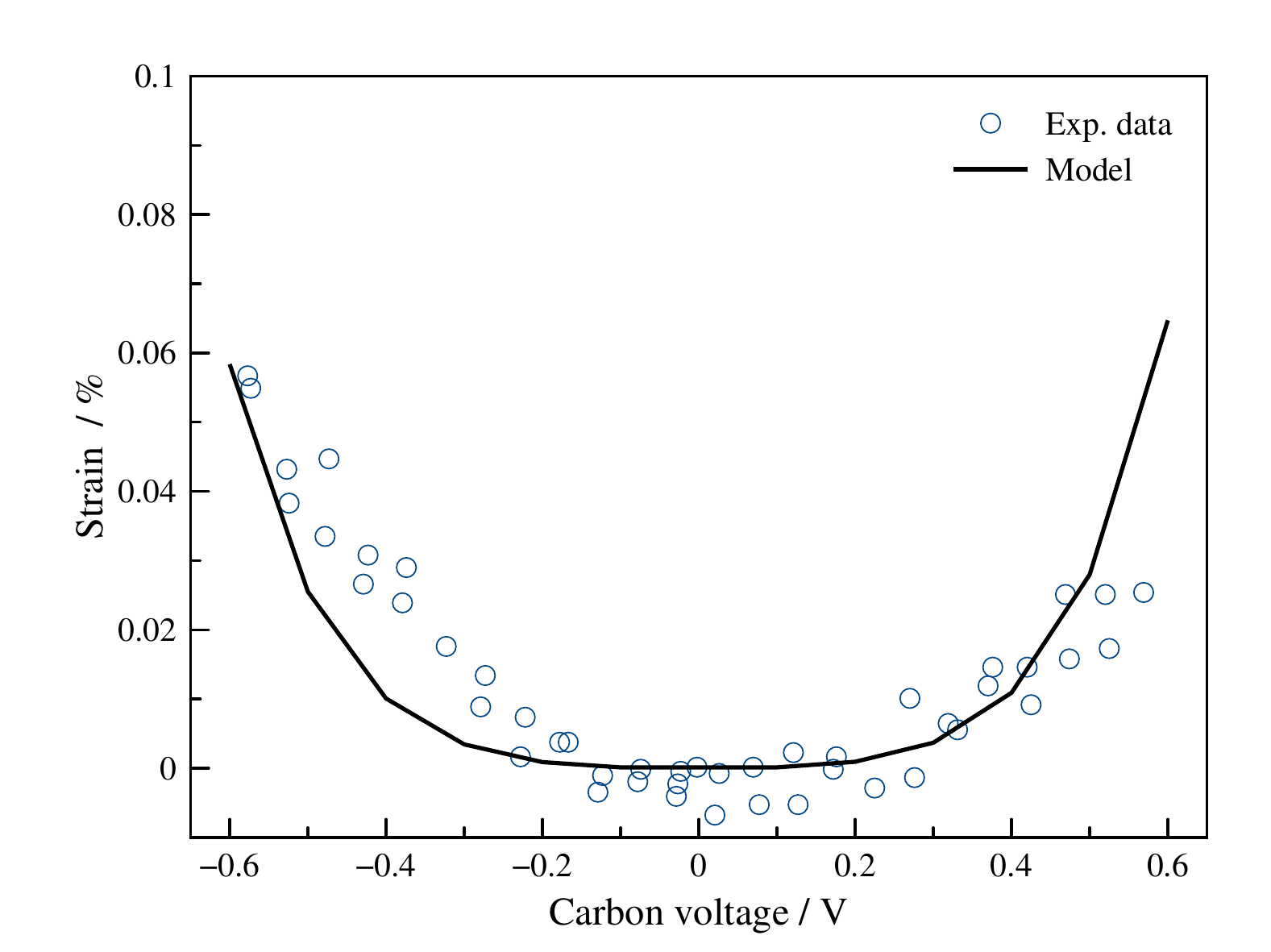} \label{fig:cscl_fit} }
\caption{Experimental\cite{koczwara2017situ} and theoretical electrosortpion-induced strains for both LiCl and CsCl aqueous solutions. The modeling results correspond to the rod diameter equal to \SI{6.5}{nm}.The comparison is made with the MC sample from Ref.\cite{koczwara2017situ}.}
\label{fig:densities}
\end{figure} 

We fit the experimentally measured strain curves\cite{koczwara2017situ} with Eq.(\ref{eq:M}) using the pore-load modulus as a free parameter. The calculations were performed for the three chosen rod diameters and the results are summarized in the Table~\ref{tab:modulus}. In the case of LiCl aqueous solution we obtain highly non-symmetric force, see Fig.(\ref{fig:licl_fit}). This is in disagreement with the experimental data and leads to the discrepancy between the obtained elastic moduli (see Table \ref{tab:modulus}). The ion specificity is taken into account only via different effective hydrated diameters of the ions\cite{nightingale1959phenomenological} (\SI{0.764}{nm} for Li, \SI{0.658}{nm} for Cs and \SI{0.664}{nm} for Cl) and, thus, leads to different shapes of the theoretical strain-voltage results. However, the magnitudes of the obtained pore-load moduli for \SI{6}{nm} and \SI{6.5}{nm} are in range with the previously reported ones\cite{ludescher2021adsorption} obtained for a similar material by analysing the adsorption-induced strain isotherms (\SI{0.32}{GPa} - \SI{2.36}{GPa}). 

\begin{table}[] 
\caption{The estimated pore-load modulus of the mesoporous structure in the nonactivated carbon sample (MC)\cite{koczwara2017situ}.}\label{tab:modulus}
\begin{tabular}{l|l|l|l}
\hline
 & 2R = \SI{6}{nm} & 2R = \SI{6.5}{nm} & 2R = \SI{7}{nm} \\ \hline
\textbf{LiCl} & \SI{0.63}{GPa}  & \SI{4.36}{GPa}   & \SI{20.88}{GPa} \\ \hline
\textbf{CsCl} & \SI{0.22}{GPa}  & \SI{1.42}{GPa}   & \SI{9.86}{GPa} \\ \hline
\end{tabular}
\end{table}

Another interesting comment can be given about the additivity of the inter-rod interactions. We performed additional calculations for similar systems but consisting only of two carbon rods immersed in the LiCl/CsCl aqueous solutions. The obtained force magnitude ($f_0$) was used to predict the generated in the seven-rod system (see, Fig.\ref{fig:system}). We assumed that the total resultant force is a vector sum of the forces generated by the nearest neighbor rods only. Thus, the direction of the resultant force is radial from the central rod and has an absolute value equal to 2$f_0$. The comparison between the additive estimated force and the one obtained for the whole seven-rod system for the case of the \SI{6.5}{nm} rod diameter is given in the Supplementary material, see Fig.S1 (a). The results agree quite well with each other and, thus, at least to some extent, the additive method can be used in this system for force estimations. It can be used to predict on the deformation for a bigger system, for example, for the one shown in Fig.S1 (b). The symmetry allows us to conclude that the strain-generated force is applied only to the "surface" rods.

\begin{figure}[ht]
\includegraphics[width=.6\textwidth]{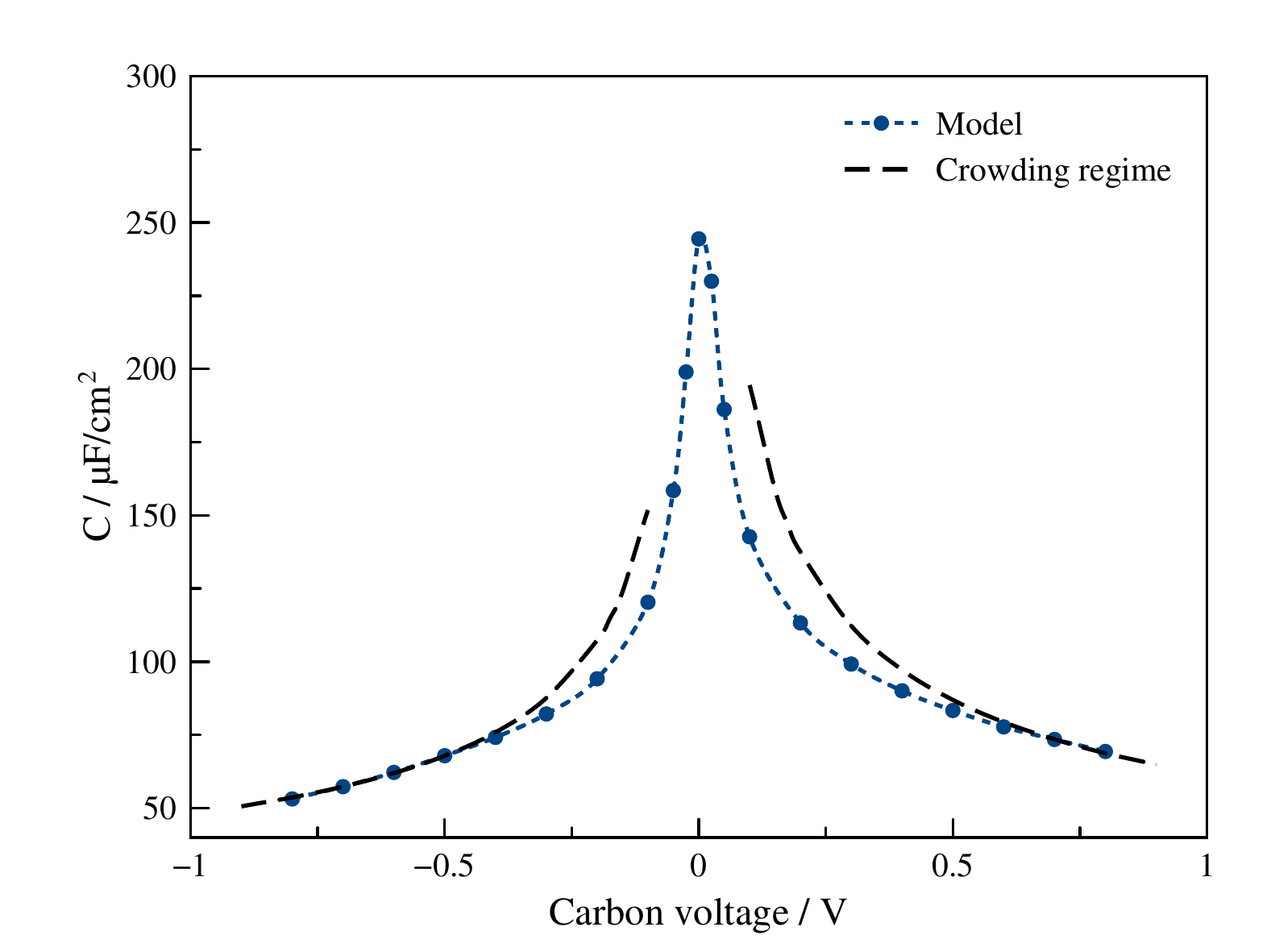} 
\caption{Differential capacitance as a function of applied voltage on the carbon rods for a LiCl aqueous solution. The dotted lines represent the crowding\cite{kornyshev2007double,fedorov2014ionic,budkov2021electric} regime, i.e. inverse square root of the voltage absolute value. } \label{fig:cap_licl} 
\end{figure} 

It is also instructive to study the differential capacitance (C) within the current model. The surface charge density is defined as the total charge on the carbon rods divided by their surface area. We performed calculations for both solutions with a fixed rod diameter, namely \SI{6.5}{nm}. The differential capacitance profiles have bell-shape\cite{kornyshev2007double} form Figs. (\ref{fig:cap_licl} and S2) and the magnitudes typical of the models of this type\cite{budkov2021electric}. The latter usually highly overestimate the real experimental values. To make the differential capacitance values more realistic, it is necessary to take into account the short-range specific interactions~\cite{goodwin2017mean,budkov2018theory} and ionic correlations~\cite{bazant2011double}. Additionally, the differential capacitance in the CMK-3 like system demonstrates existence of the crowding regime (the differential capacitance is inversely proportional to the square root of the voltage)~\cite{kornyshev2007double,fedorov2014ionic,budkov2021electric} observed at sufficiently high voltages $\approx \pm$ \SI{0.5}{\volt}. As in the case of a flat EDL~\cite{maggs2016general}, the ionic size asymmetry effect makes the differential capacitance slightly asymmetric, the latter is obviously more pronounced for the LiCl case. 

\section{Concluding remarks}
In conclusion, we would like to summarize the main results of our work. Within the simple mean-field model we estimated the inter-rod force generated in an CMK-3-like structure caused by ion osmotic pressure and electrostatic interactions. Using the previously published information about the material structure~\cite{koczwara2017situ} we estimated the pore-load modulus based on electrosorption experimental data and found an agreement with the data obtained for the similar material by conventional adsorption measurements. Thus, the calculated force magnitude has a realistic order indicating the potential of our electrosorption-induced deformation model (and confirming the discussions of deformation mechanism in the Ref.\cite{koczwara2017situ}), as well as the previously suggested~\cite{kolesnikov2020density} material deformation mechanism. We demonstrated that the resulting force acting on each carbon rod can be approximated on an additive (pair-wise) basis, which can be used further to upscale the model to describe the macroscopic sample deformation. Finally, we studied the differential capacitance in the same pore geometry and demonstrated the crowding regime at sufficiently high applied voltages. We believe that our results can be useful for the interpretation of electrosorption/adsorption-induced deformation data for a CMK-3-like materials as well as for others with non-convex pore geometry.

\begin{acknowledgements}
YAB thanks the Russian Science Foundation (Grant No. 21-11-00031). The work of DAM was financed within the Project Teams framework of MIEM HSE. In part, the calculations were performed on the supercomputer facilities provided by NRU HSE. This research was supported in part through the computational resources of the HPC facilities at HSE University\cite{hse_cluster}. Additionally the computations for this work were done using resources of the Leipzig University Computing Centre.
\end{acknowledgements}



\section*{Supporting information}
\begin{figure}[ht]
\subfloat[]{\includegraphics[width=.45\textwidth]{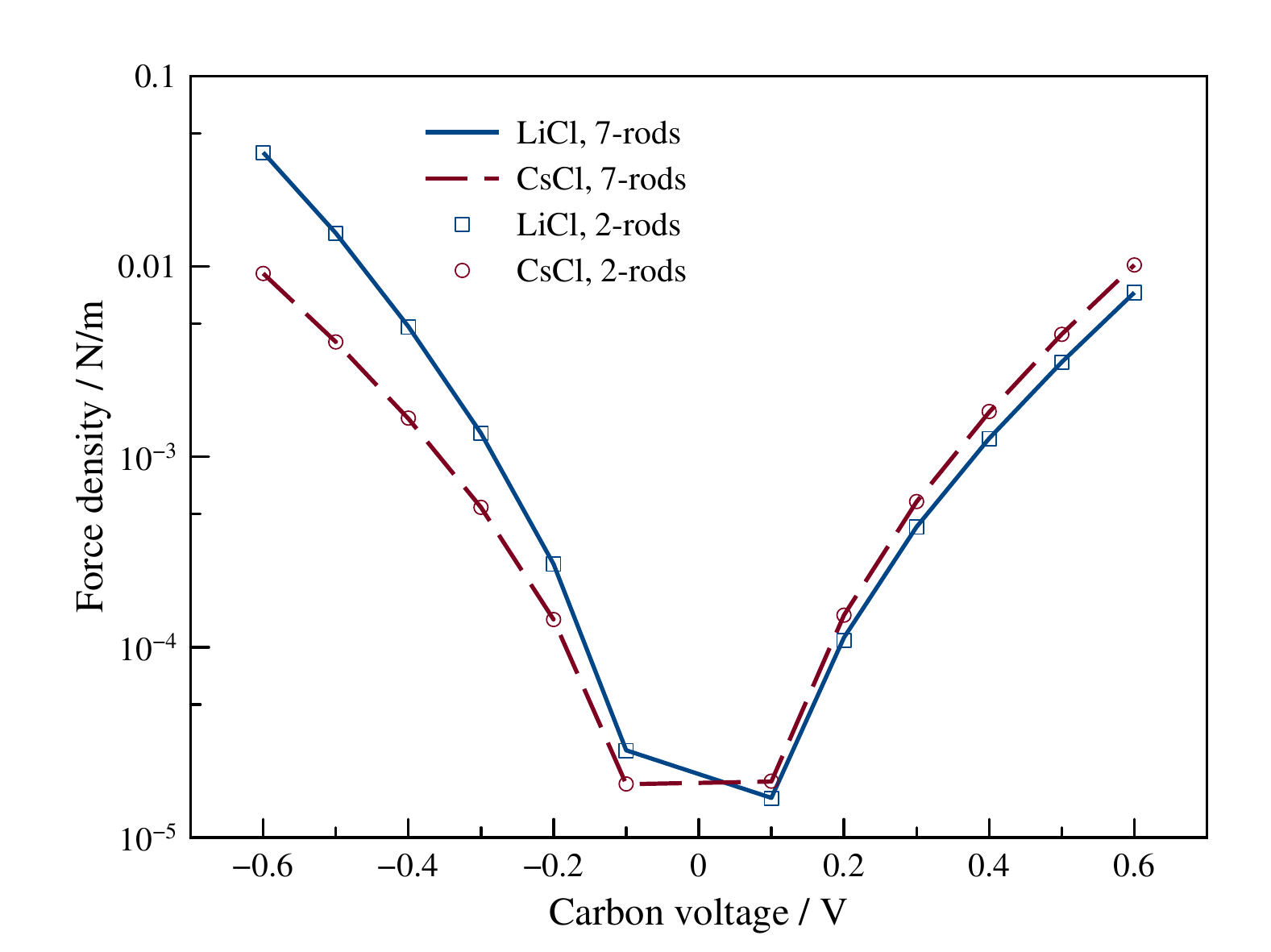} \label{fig:force_comparison}  } 
\subfloat[]{\includegraphics[width=.35\textwidth]{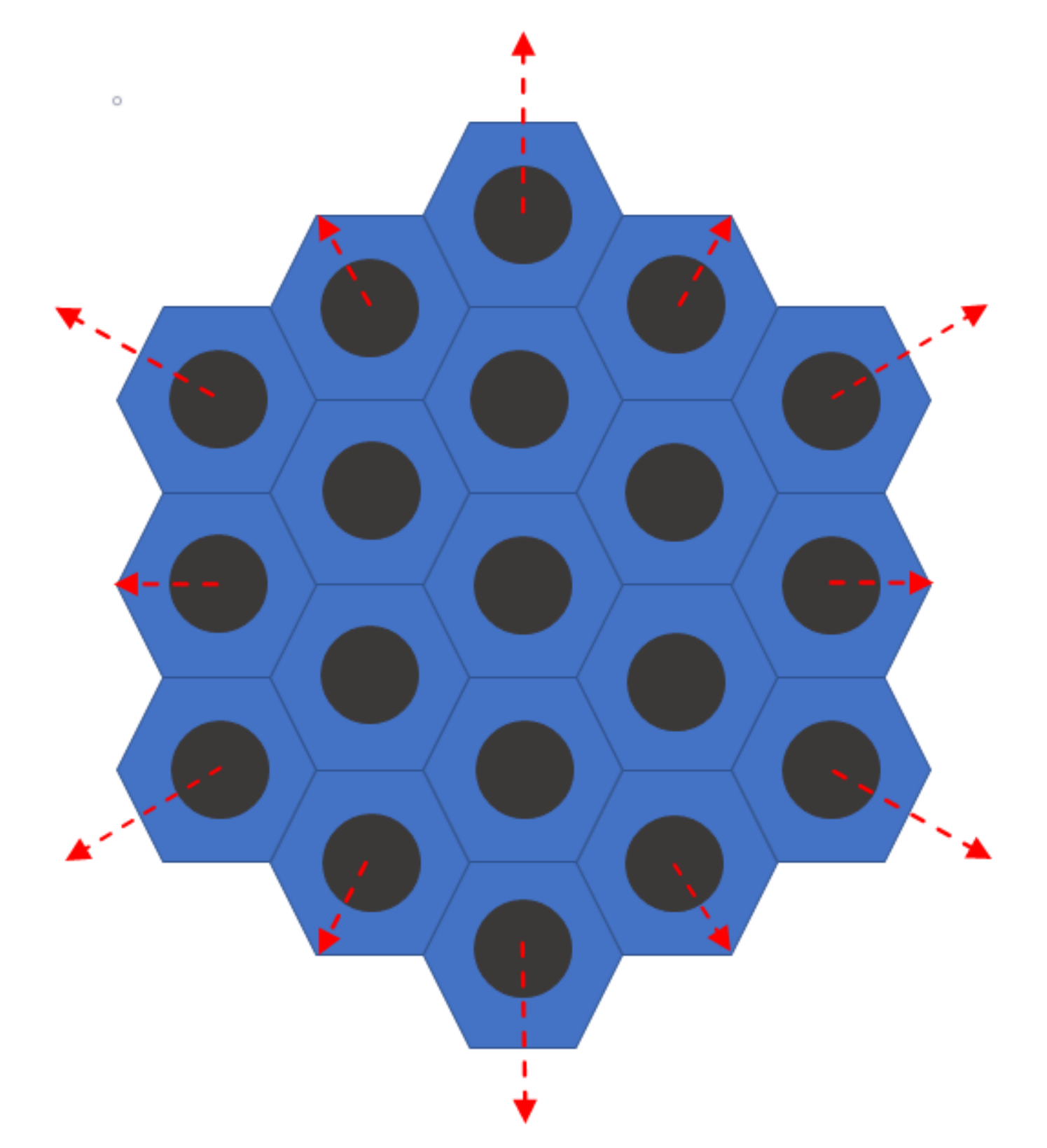} \label{fig:3_layers} }
\caption{(a) Comparison of the linear force density (the resultant force per unit rod length) calculated in the 7-rod system and within additive approximation (2-rods) for LiCl and CsCl solutions with the rod diameter equal to \SI{6.5}{nm}. (b) Schematic distribution of the electrosorption-induced forces in additive approximation.}
\label{fig:densities}
\end{figure} 

\begin{figure}[h]
\includegraphics[width=.6\textwidth]{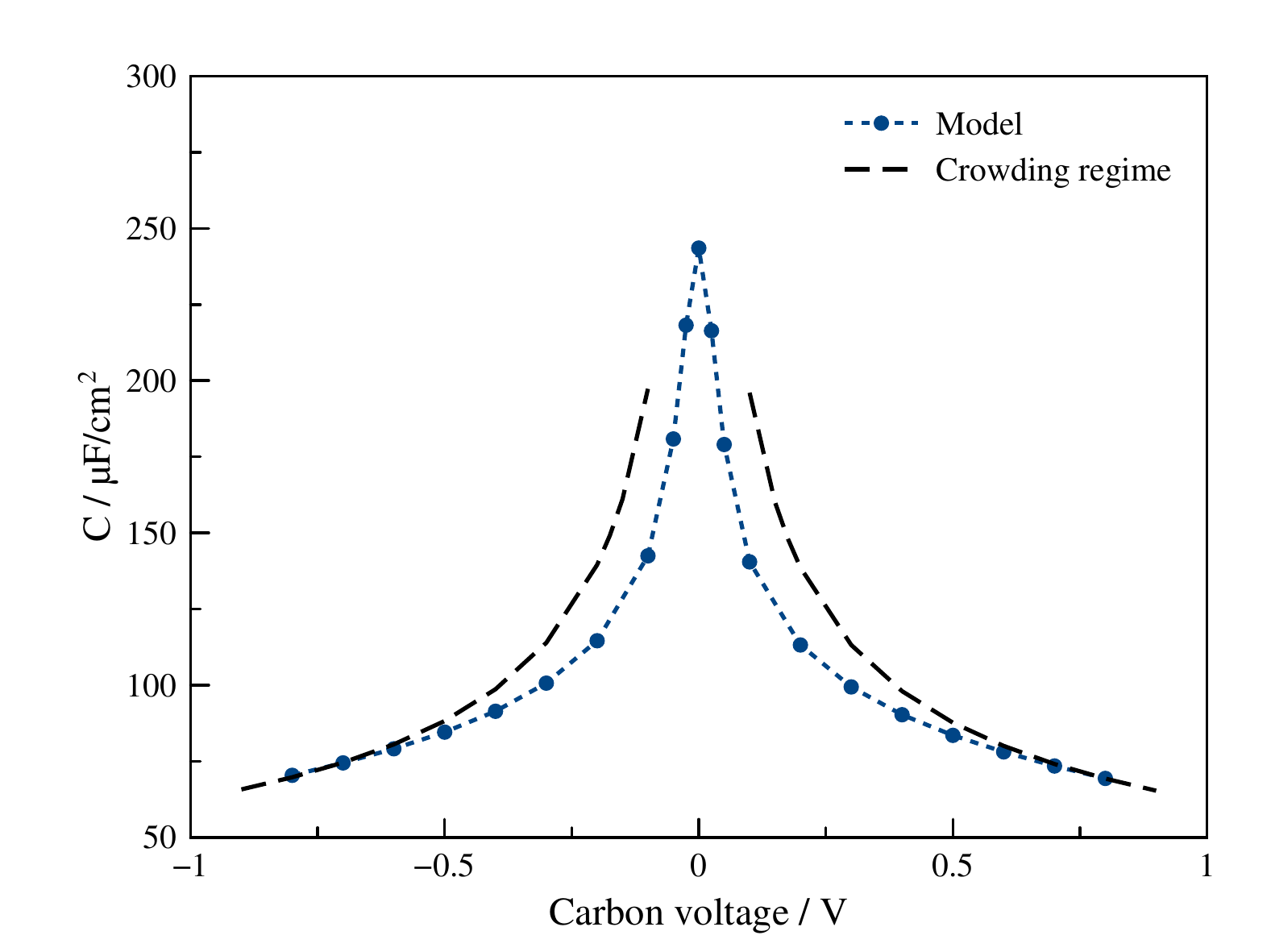} 
\caption{Differential capacitance as a function of applied voltage on the carbon rods for the CsCl aqueous solution. The dotted lines represent the crowding~\cite{kornyshev2007double,fedorov2014ionic,budkov2021electric} regime, i.e. inverse square root of voltage absolute value. }  
\end{figure} 

\bibliography{lit}

\end{document}